\documentclass[conference]{IEEEtran}

\RequirePackage[l2tabu, orthodox]{nag}

\usepackage{cite}
\usepackage{amsmath,amssymb,amsfonts}
\usepackage{algorithmic}
\usepackage{threeparttable}
\usepackage[pdftex]{graphicx}
\graphicspath{{./figures/}{./resources/}}
\DeclareGraphicsExtensions{.pdf,.png,.jpg,.jpeg}

\usepackage{tabularx}
\usepackage{textcomp}
\usepackage{xcolor}
\usepackage[normalem]{ulem}
\usepackage{multicol}
\usepackage{multirow}

\ifCLASSOPTIONcompsoc
\usepackage[caption=false, font=normalsize, labelfont=sf, textfont=sf]{subfig}
\else
\usepackage[caption=false, font=footnotesize]{subfig}
\fi

\usepackage{booktabs, makecell}
\newcommand{\ra}[1]{\renewcommand{\arraystretch}{#1}}

\usepackage{xspace}
\makeatletter
\DeclareRobustCommand\onedot{\futurelet\@let@token\@onedot}
\def\@onedot{\ifx\@let@token.\else.\null\fi\xspace}

\makeatother

\usepackage{hyperref}

\begin{document}

\title{Graph Neural Networks Based Anomalous RSSI Detection}


\author{\IEEEauthorblockN{%
    Bla\v{z}~Bertalani\v{c}\IEEEauthorrefmark{1}, Matej Vnučec\IEEEauthorrefmark{1}, and Carolina Fortuna\IEEEauthorrefmark{1}
}\IEEEauthorblockA{%
    \IEEEauthorrefmark{1}Department of Communication Systems, Jo\v{z}ef Stefan Institute, Slovenia\\
}
\{blaz.bertalanic, carolina.fortuna\}@ijs.si%
\\
}

\maketitle

\begin{abstract}
In today's world, modern infrastructures are being equipped with information and communication technologies to create large IoT networks. 
It is essential to monitor these networks to ensure smooth operations by detecting and correcting link failures or abnormal network behaviour proactively, which can otherwise cause interruptions in business operations.
This paper presents a novel method for detecting anomalies in wireless links using graph neural networks. The proposed approach involves converting time series data into graphs and training a new graph neural network architecture based on graph attention networks that successfully detects anomalies at the level of individual measurements of the time series data. The model provides competitive results compared to the state of the art while being computationally more efficient with $\approx$171 times fewer trainable parameters.

\end{abstract}


\section{Introduction}

The Internet of Things (IoT) has become one of the most important concepts in modern wireless networks. The availability of low-cost sensors with connectivity capabilities allows us to augment existing infrastructures and processes, such as transportation or power delivery networks, with new information that enables effective monitoring. However, each IoT sensor deployed increases network complexity and introduces new challenges to solve. Traditionally, wireless connectivity issues were monitored manually using a set of predefined metrics that determined the potential presence of anomalies and their nature. Furthermore, with the transition to fifth generation mobile networks (5G) and beyond, the capabilities of the technology such as higher peak data speeds, ultra-low latency, more reliability, massive network capacity, and higher availability, will lead to increasingly complex IoT networks, impossible to be manually monitored. To ensure the reliability of such complex IoT networks, new automated solutions must be developed \cite{annaswamy2016emerging}. 
 
 According to \cite{cerar2020anomaly}, when a monitoring infrastructure consisting of multiple devices is operated over a long period of time, there is a possibility of malfunctions and downtime caused by various factors such as software bugs, physical damage, transceiver degradation, and line of sight obstacles. The authors of \cite{cerar2020anomaly} have identified four different types of anomalies that can occur in the link layer of wireless communications, namely Sudden Link Degradation (SuddenD), Sudden Link Degradation with Recovery (SuddenR), Instantaneous Link Degradation (InstaD), and Slow Link Degradation (SlowD). These anomalies are represented in Fig.~\ref{fig:representation}. 

\begin{figure}[thb]
        \label{fig:representation}
	\centering
	\subfloat[SuddenD anomaly\label{fig:synthetic:step:norecovery}]{
		\includegraphics[width=.47\linewidth]{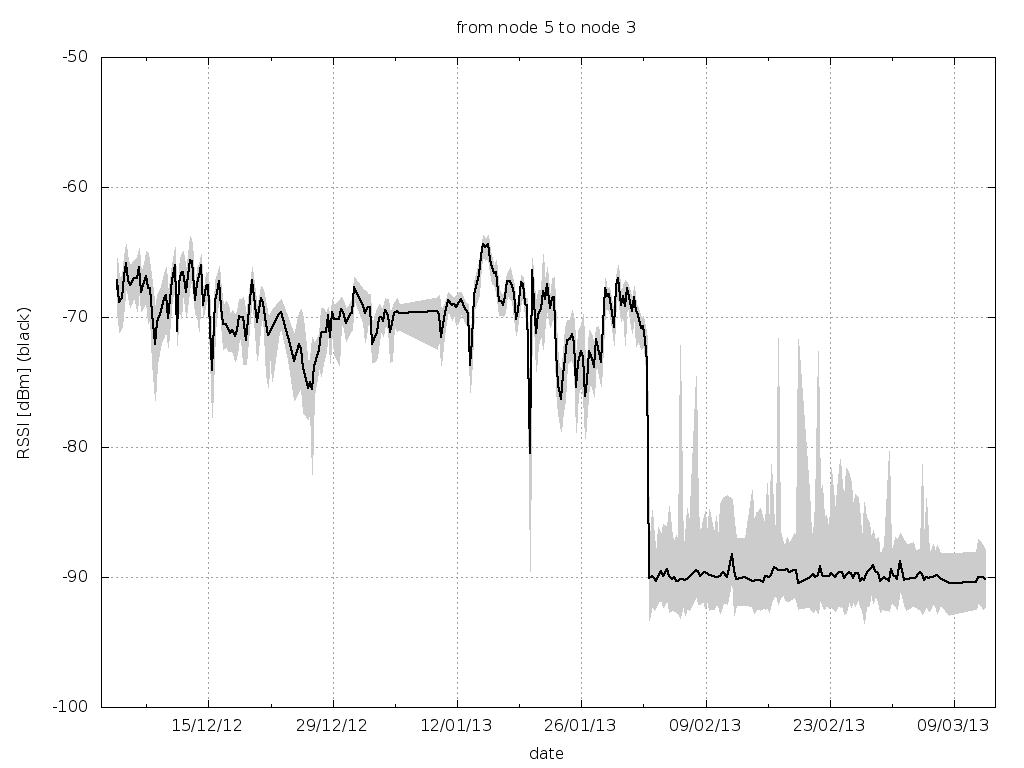}
	}
	\subfloat[SuddenR anomaly\label{fig:synthetic:step:recovery}]{
		\includegraphics[width=.47\linewidth]{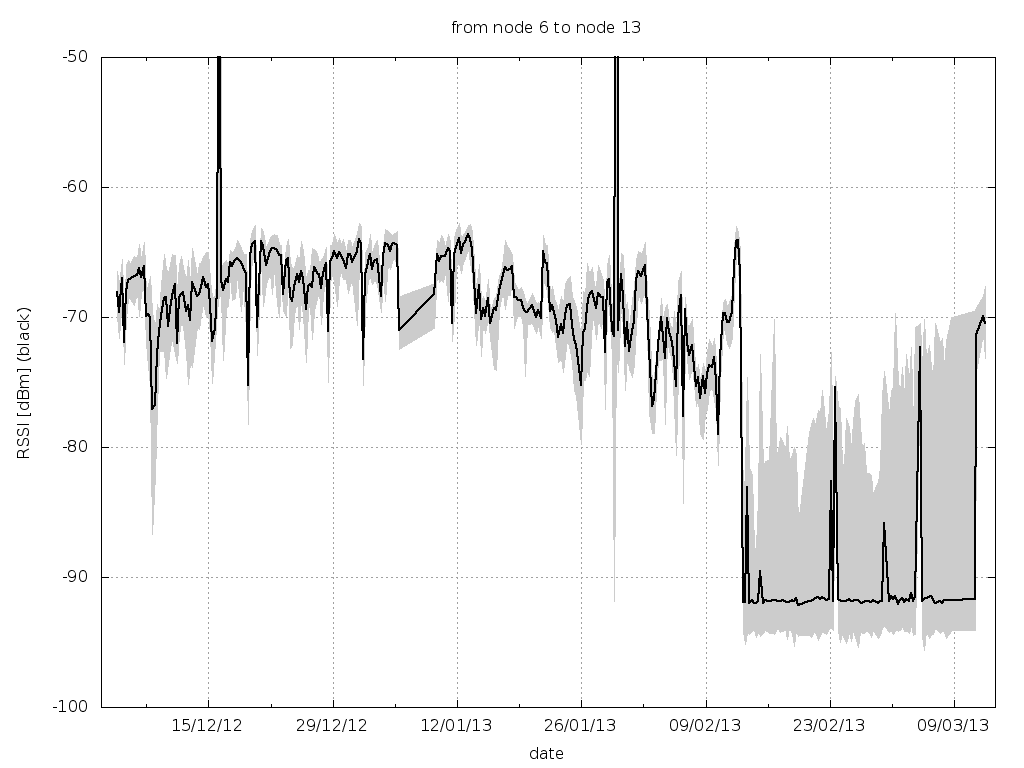}
	}
	
	\subfloat[InstaD anomaly\label{fig:synthetic:step:spikes}]{
		\includegraphics[width=.47\linewidth]{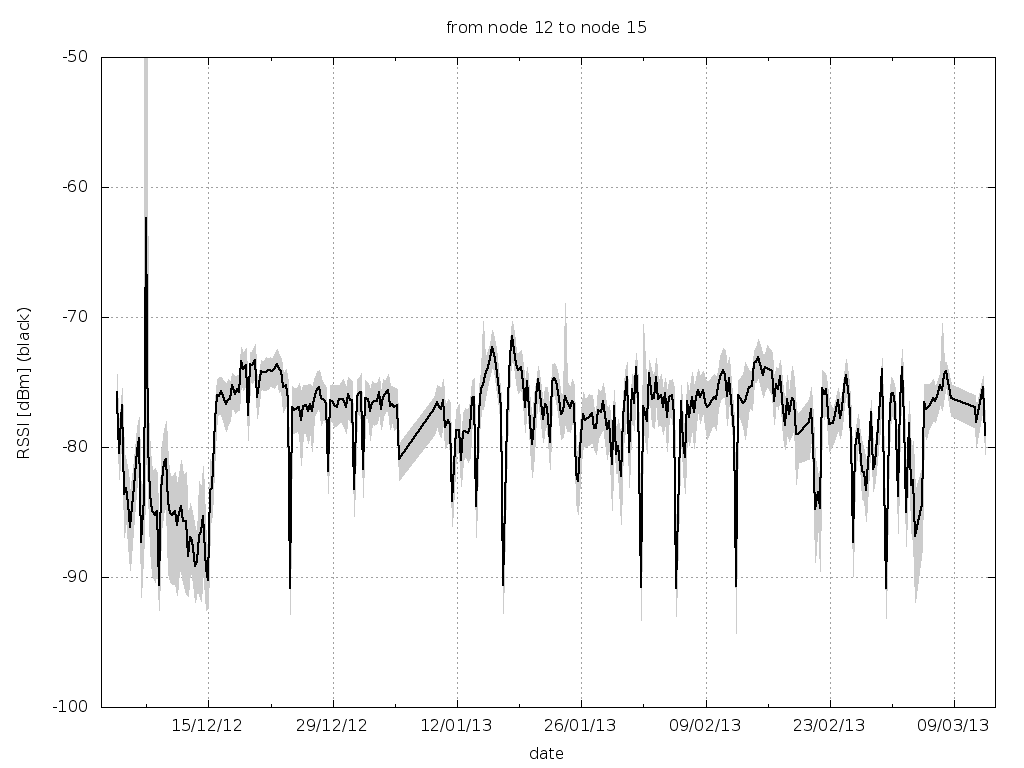}
	}
	\subfloat[SlowD anomaly\label{fig:synthetic:step:slow}]{
		\includegraphics[width=.47\linewidth]{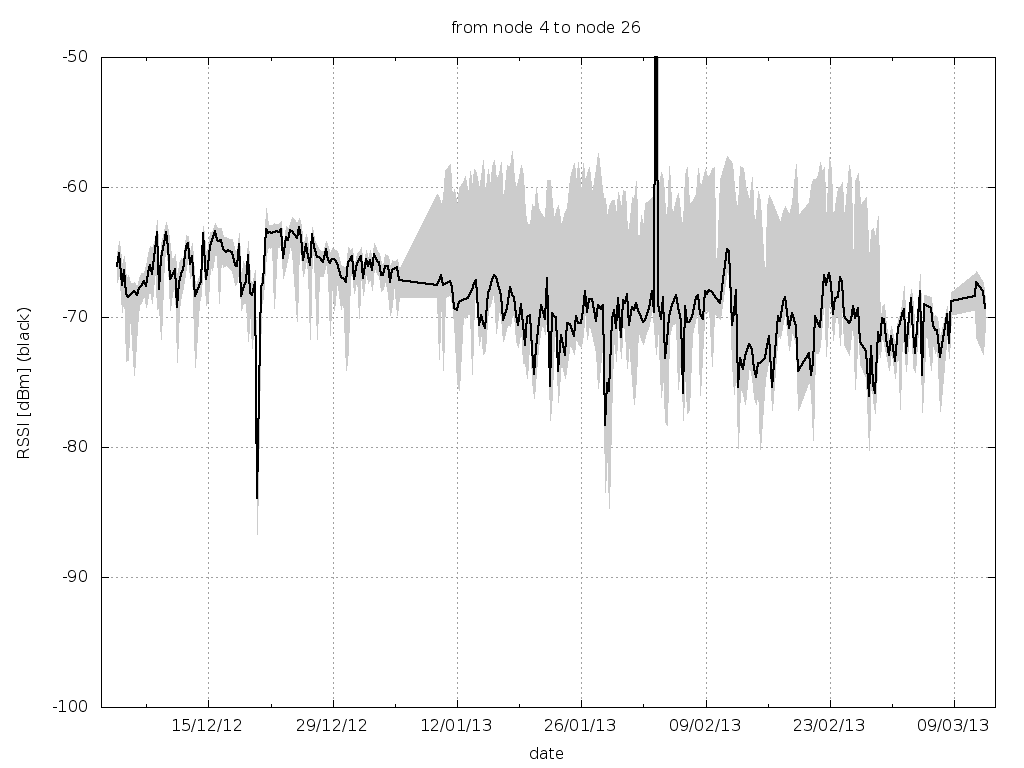}
	}
	\caption{Time-series representation of the links and of the four types of anomalies considered in~\cite{cerar2020anomaly}.}
	\label{fig:representation}
\end{figure}

Numerous automated solutions have been proposed to facilitate the effective management of large IoT networks. These solutions include network monitoring~\cite{silva2019m4dn}, malfunction detection~\cite{sheth2006mojo}, and specific anomaly classification~\cite{cerar2020anomaly,bertalanivc2022resource}. Typically, machine learning algorithms are used to develop such automated solutions. But the limitation of these methods is that they can only detect and classify an anomaly or malfunction, but then demand additional manual inspection of network signals to determine the severity of the malfunction and appropriate steps to mitigate the problem. Another problem of such systems is that they can only detect problems within a fixed time window and are unable to precisely pinpoint the time step at which an anomaly occurred within the link.

Graph neural networks (GNNs)\cite{wu2020comprehensive} have recently gained popularity for various applications in non-Euclidean spaces, although they are not limited to this type of space\cite{zhou2020graph}. Researchers from various fields have used GNNs to address a wide range of problems, one of which is an adaptation for time series (TS) anomaly detection tasks, particularly for multivariate TS~\cite{ScheinertGNNanomalyCloud, deng2021graph}. GNNs can be used to model the temporal relationships between data points in the time series, as well as model complex interactions and relationships between the features of each data point. Recently, a novel method on transforming univariate TS data into graphs that can be utilised for a more in depth analysis of deviations happening within the TS, such as wireless link layer anomalies, has been introduced \cite{blazWCNC}. 

In this paper, we analyse the performance of graph neural networks, for detecting and localising wireless link anomalies. The contributions of this paper are:

\begin{itemize}

    \item A new approach for wireless link layer anomaly detection, that works on per point basis which enables for not only detecting an anomaly, but also localising it within the TS and determine its duration.
    
    \item A novel graph neural network model, based on Graph Attention Networks, for classifying TS points that achieves similar performance to the state of the art, while having an $\approx$171 times less weights.
    
\end{itemize}

This paper is organised as follows. We discuss the related work in Section~\ref{sec:related} while Section~\ref{sec:method} elaborates on the proposed method, transformation and architecture. Section~\ref{sec:methodology} presents the methodological aspects of the work while Section~\ref{sec:evaluation} provides the evaluation. Finally, Section~\ref{sec:conclusion} concludes the paper.

\section{Related work}
\label{sec:related}

To support our contributions and put our work in perspective, we first analyse related work with respect to graph neural networks for time series analysis and then narrow down to works for anomaly detection in wireless communication networks.  

\subsection{Graph neural networks on time series data}
The majority of the current state-of-the-art (SotA) research in the field of time series analysis focuses on multivariate time series data forecasting. In contrast, research on the classification of time series data is limited.

Cao~\textit{et. al.}\cite{cao2020spectral} were among the pioneers in proposing the use of graph representation of multivariate time series to solve forecasting problems. Their approach involved transforming the time series into a graph by training a Gated Recurrent Unit (GRU) layer and using an attention mechanism to construct a weight matrix that served as the adjacency matrix for the graph representation. This learned graph was then used in a spectral graph neural network for forecasting. Similar a graph structure learning layer approach was proposed by \cite{wu2020connecting}, also for the purpose of forecasting multivariate TS.

Researchers also showed the potential of utilising the GNNs for multivariate TS anomaly detection, but their approach was again based on forecasting methods. For example, Deng~\textit{et. al.}~\cite{deng2021graph} proposed GNNs to forecast multivariate sensor signals based on GAT and if there is a deviation between the forecast and actual behaviour they label that as an anomaly. Similar work was also presented by \cite{ScheinertGNNanomalyCloud} to try to detect anomalies in cloud infrastructure monitoring. Both studies work on per TS data point analysis, but only utilise a simple thresholding method rather than the classification power of GNNs to do so.

More recently, there have been initial attempts to classify univariate TS data. For instance, \cite{Xuan_9695244} proposed an approach that combined a visibility graph trained concurrently with a radio signal modulation classification model. Similarly, \cite{Xiu_9778887} utilized a time-labelled visibility graph as a component of their electrocardiogram classification model. Another novel approach for transforming TS data into graphs was proposed by \cite{blazWCNC}, where they created a graph based on an adjacency matrix with Markov Transition Fields, which was later used to classify wireless anomalies. But similar to multivariate TS analysis with graphs, in univariate analysis the researchers analysed the TS graphs as a whole rather than at the granular level of nodes representing a point within a TS.

\subsection{Wireless network anomaly detection}

Current research on anomaly detection in wireless networks mainly focuses on intrusion, fraud, fault detection, IoT event detection, system health monitoring, and natural disasters~\cite{chandola2009anomaly}. There are many works focusing on wireless anomaly detection that are based on the classical machine learning algorithms. Salem~\cite{salem2014anomaly} proposed the use of SVM, decision trees, logistic regression, Naive Bayes, and decision tables to discriminate between anomalies in medical data gathered from IoT networks. Wazid~\cite{wazid2016efficient} proposed an unsupervised k-means based way of detecting intrusions in wireless traffic data improved by~\cite{ahmad2019hybrid} using the k-medoid algorithm. In our recent work~\cite{cerar2020anomaly}, we detected four link layer anomalies based on unsupervised machine learning approaches (LOF, Isolation Forests, and one-class SVM) and supervised learning (Logistic regression, random forest, and SVM). This was further improved by us with the use of supervised deep learning approaches~\cite{bertalanic_9606264, bertalanivc2022resource}. 

In general, unsupervised learning using autoencoders (AE) is the most commonly used deep learning approach for anomaly detection. Thing\cite{thing2017ieee} conducted an evaluation of four different AE models for intrusion detection using IEEE 802.11 network data . Similarly, Ran et al.\cite{ran2019semi} employed a semi-supervised AE model to detect four different types of cyber attacks in captured IEEE 802.11 network data. 

\section{The proposed method}
\label{sec:method}

Suppose a large IoT network is deployed in a smart infrastructure and we expect an uninterrupted data communication to ensure uninterrupted business process. To accomplish this, an automatic monitoring system of such IoT network is designed to quickly detect and mitigate potential problems and anomalies. Devices within the IoT network produce time series which is used for monitoring. We formulate the anomaly detection task in the system as a classification problem. To analyse the input data in graph form, the input time series $S$ needs to be transformed into graphs using a transformation function $T$. Such transformed TS can then be fed to the function $\Phi$, that represents a GNN, which maps it to a vector of target classes $P_{1\times N} = \vec{p_N}= {p_1, p_2,..., p_N}$, where $N$ represents the length of the time series data, as provided in Eq.~(\ref{mathref:classification}).

\begin{equation}
	\label{mathref:classification}
	P_{1\times N}=\Phi(G) = \Phi(T(S))
\end{equation}

We consider our problem as a binary classification problem where the set of target classes is $P=\{anomalous, non-anomalous\}$. Each node, that represents a point within a TS, is then classified either as anomalous or non-anomalous, producing a binary vector $P_{1\times N}$.

\begin{figure}[htb]
	\centering
	\includegraphics[width=\linewidth]{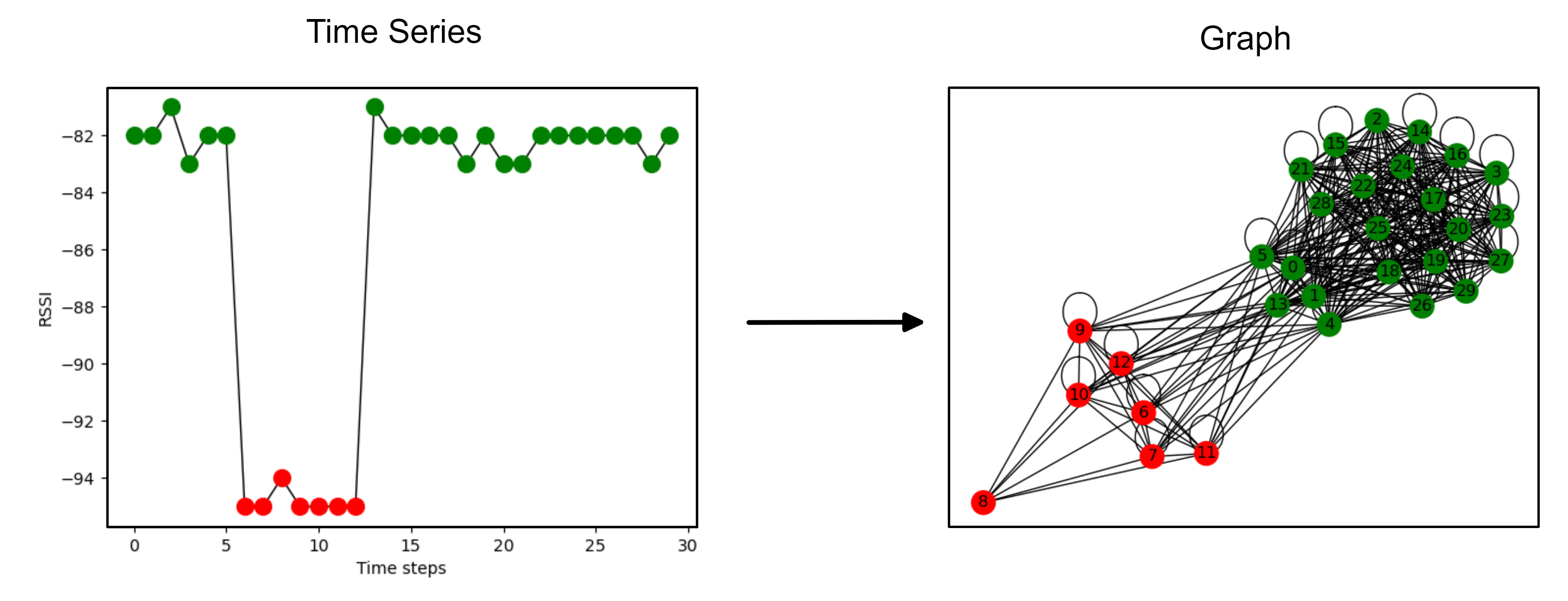}
	\caption{A simple example of TS to graph transformation. Non-anomalous nodes are coloured in green, while the anomalous are coloured red.}
	\label{fig:TS2graph}
\end{figure}

\subsection{Time-series to graph transformation}
\label{sec:graph_transformation}
We define the transformation $T$ depicted in Fig.~\ref{fig:TS2graph} that transforms the input TS $S$ to the graph representation $G$ as in Eq. \ref{mathref:transformation}. The time-series to graph transformation is realized in three steps: node determination, adjacency matrix computation, and final graph representation. We decided to adopt Markov Transition Field (MTF)~\cite{wang2015encoding} representation of TS as an adjacency matrix for the transformation $G$. The MTF representation is a sparse matrix of transition probabilities between the values in the TS and can be interpreted as weighted edges of a graph.


\begin{equation}
	\label{mathref:transformation}
	G=T(S)
\end{equation}

To create the graph $G$, the first step is to convert each point in a time series $S_{1\times N} = \vec{s_N}= {s_1, s_2,..., s_N}$ into a node in $G$, with the value at that point becoming the node's label (feature). The number of nodes in $G$ is equal to the length of the time series.

Next, the adjacency matrix $A$ corresponding to $G$ is determined using the Markov Transition Field (MTF) technique. This involves quantizing the values of the time series into a finite set of bins, and then computing the probability of transitioning from one bin to another. This preserves the temporal information of the time series. MTF matrix is defined as follows:

\begin{equation}
	\label{mathref:MTF}
	A =  
	\begin{pmatrix}
		w_{ij|s_1 \in r_i,s_1 \in r_j} & ... &  w_{ij|s_1 \in r_i,s_N \in r_j}\\
		w_{ij|s_2 \in r_i,s_1 \in r_j} &  ... &  w_{ij|s_2 \in r_i,s_N \in r_j}\\
          \vdots &  \ddots & \vdots\\
		w_{ij|s_N \in r_i,s_1 \in r_j} & ... &  w_{ij|s_N \in r_i,s_N \in r_j}\\
	\end{pmatrix}
\end{equation}


MTF encodes multi-span transition probabilities of the time series by assigning probabilities from quantiles at one time step to those at another time step, and representing them as $w_{ij}$. 

The third and final step involves constructing the graph $G$ from the transition matrix $A$. The edges of $G$ are created based on $A$, where any $w_{ij} > 0$ generates an edge between nodes $i$ and $j$ with $w_{ij}$ serving as the edge weight. Consequently, each resulting graph $G$ consists of a set of nodes N that represent each time step in the time series with its value, a set of edge indexes, and a set of edge weights. 

An example of this three step process is depicted in  Fig.~\ref{fig:TS2graph}, where a TS $S_{1 \times 30}$ is transformed by the $T$ transformation into a graph $G$ with 30 nodes connected by 493 edges. The nodes representing the anomalous part of the wireless link are red, while the non-anomalous nodes are green. As it can be observed in Fig.~\ref{fig:TS2graph}, nodes that are most similar to each other group together and have more connections between themselves.

\subsection{Proposed GNN architecture}
\label{sec:model}

\begin{figure}[htb]
	\centering
	\includegraphics[width=0.5\linewidth, angle=90]{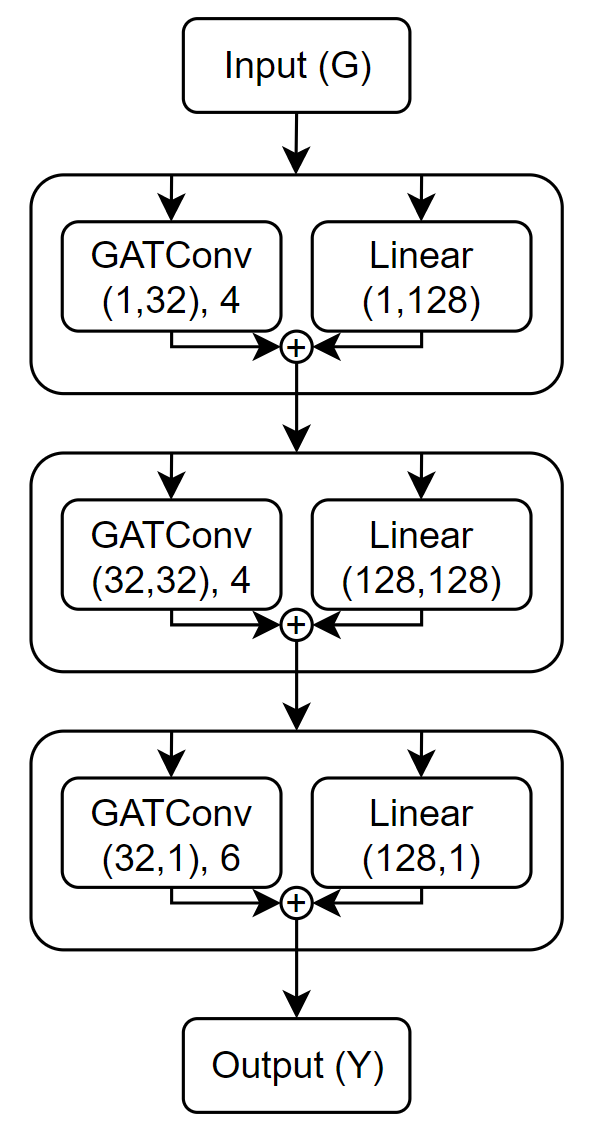}
	\caption{Proposed GNN architecture.}
	\label{fig:model}
\end{figure}

We define our prediction model as function $\Phi$ that transforms the input transformed data $G$ to the set of target classes $P_{1\times N}$ as provided in Eq.~\ref{mathref:classification} and depicted in Figure \ref{fig:model}. 

We have designed our GNN architecture inspired by the the Graph Attention Network (GAT)~\cite{velickovic2018graph}. GAT improves upon classical Graph Neural Networks by applying self-attention over the node features, significantly improving classification performance for node classification tasks across different domains. The final model architecture was selected based on empirical testing, where as a metric we were considering the best ratio between number of weights and performance of the model.

The proposed architecture is realised with 3 GAT layers that are combined with Linear layers. Linear layers work as learnable skip connections~\cite{skip_connection}, that improve training and convergence of the model. In our model all Linear layers have 128 nodes. All GAT layers consists of 32 filters, they only differentiate by the number of self-attention heads. First two layers consists of 4 self-attention heads, while the third employs 6 heads. The number of layers, filters, and heads were selected with grid search process, optimising for the best performance of the model. The output of final layer is finally sent to the output layer Y, that produces a vector $P_{1\times 300}$ of size 300 representing 300 nodes, or measurements, within the TS. For each of the 300 nodes network returns probability of it being Anomalous or Non-anomalous. All hidden layers use the ReLU function for activation except for the final layer that uses a Sigmoid function. 

\begin{table}[htbp]
	\centering
	\ra{1.2}
	\begin{threeparttable}[b]
		\caption{Synthetic anomaly injection method similar to~\cite{cerar2020anomaly}.}
		\label{tab:injection-scenario}
		\begin{tabularx}{\linewidth}{@{}lllll@{}}
			\toprule
			Type
			& Links
			& Affected
			& Appearance
			& Persistence
			\\\midrule
			
			SuddenD 
			& \multirow{4}{*}{2\,123}
			& \multirow{4}{*}{33\% (700)}
			& once, [200$^\textrm{th}$,~280$^\textrm{th}$] 
			& for $\infty$
			\\
			
			SuddenR 
			&
			&
			& once, [25$^\textrm{th}$,~275$^\textrm{th}$]
			& for [5,~20]
			\\
			
			InstaD 
			&
			&
			& on $\approx$1\% of a link
			& for 1 datapoint
			\\
			
			SlowD 
			&
			&
			& once, [1$^\textrm{st}$,~20$^\textrm{th}$]
			& for [150, 180]$^\dagger$
			\\
			
			\bottomrule
		\end{tabularx}
		\begin{tablenotes}
			\item[$\dagger$] RSSI$(x, \textrm{start})$ $\leftarrow$ RSSI$(x)$ + $\min(0, -\textrm{rand}(0.5, 1.5)\cdot(x-\textrm{start}))$
		\end{tablenotes}	
	\end{threeparttable}
\end{table}
\begin{table*}[htbp]
	\centering
	\ra{1.2}
	\footnotesize
	\begin{threeparttable}[b]
		\caption{Classification results of proposed model compared to the SoTA.}
		\label{tab:results-classification}
		\begin{tabular}{lllllllllllllllll}
			\toprule
			
			\multirow{2}{*}{Class}
			&& \multicolumn{3}{c}{Proposed method}
			& \phantom{}
			& \multicolumn{3}{c}{RP~\cite{bertalanivc2022resource}}
			& \phantom{}
			& \multicolumn{3}{c}{GASF~\cite{bertalanivc2022resource}}
			& \phantom{}
			& \multicolumn{3}{c}{GADF~\cite{bertalanivc2022resource}}
			\\\cmidrule{3-5}\cmidrule{7-9}\cmidrule{11-13} \cmidrule{15-17}
			
			& 
			& Prec.
			& Rec.
			& F1
			& 
			& Prec.
			& Rec.
			& F1
			& 
			& Prec.
			& Rec.
			& F1
			& 
			& Prec.
			& Rec.
			& F1
			\\\midrule

		Anomalous	&& 0.93 & 0.95 & 0.94 &	& 0.99 & 1.00 & \textbf{0.99} &	& 0.97 & 0.98 & 0.98 &	& 0.91 & 0.92 & 0.91\\
			
			Non-anomalous	&& 0.98 & 0.97 & \textbf{0.98} &	& 0.99 & 0.96 & 0.97 &	& 0.94 & 0.93 & 0.94 &	& 0.97 & 0.97 & 0.97  \\
		
			\bottomrule
		\end{tabular}	
	\end{threeparttable}
\end{table*}

\begin{table}[htbp]
	\centering
	\ra{1.2}
	\begin{threeparttable}[b]
		\caption{Model trainable weights comparison.}
		\label{tab:num-weights}
		\begin{tabular}{cc}
			\toprule
			Model
			& Weights
			\\\midrule
			
			Our model
			& \textbf{$\approx$0.035M}
			\\
			
			RP/GASF/GADF model \cite{bertalanivc2022resource}
			& $\approx$6.000M
			\\

			\bottomrule
		\end{tabular}
	\end{threeparttable}
\end{table}

\section{Methodology}
\label{sec:methodology}
This section first elaborates on methodological aspects of dataset preparation followed by model training and evaluation.

\subsection{Dataset}
Rutgers WiFi dataset~\cite{rutgers-noise-20070420} was utilized as our real-world measurement dataset for the proposed GNN training. The dataset is consists of link traces obtained from 29 nodes, recorded at 5 distinct noise levels. The dataset comprises raw Received Signal Strength Indicator (RSSI) values, sequence numbers, source node ID, destination node ID, and artificial noise levels. Each RSSI value represents the signal strength of a received packet transmitted every 100 milliseconds over a 30-second period, resulting in traces that consist of 300 RSSI samples. To produce a suitable dataset for our experiments, we synthetically inject four anomalies, that were defined by~\cite{cerar2020anomaly}, according to the guidelines in Table~\ref{tab:injection-scenario}. We only considered links without packet loss. This gave us a dataset consisting of four different types of anomalous and non-anomalous links. The final dataset had 8492 samples, with 700 samples for each anomaly while the remaining 5692 samples were non-anomalous. 

Such produced traces were then transformed into graphs based on the procedure in Section~\ref{sec:graph_transformation}. Based on the empirical testing, we determined that the best number of bins for the MTF calculation is equal to the length of the TS trace, in our case that equalled to 300 bins.

\subsection{Model training and evaluation}

Model training was done using 10-fold stratified shuffle split technique. This involved shuffling and splitting the data into a training set and a test set in a ratio of 80:20. To ensure credible results the process was repeated 10 times. Due to the class imbalance, where we have significantly more non-anomalous measurements compared to 
anomalous ones, we assigned class weights during the training process which were equal to the inverse proportions of number of samples for each class.

The trained models performance was evaluated using standard classification matrices Precision, Recall, and F1-score. The F1-score is expressed as a harmonic mean between $\textrm{Precision} = \frac{\textrm{TP}}{\textrm{TP} + \textrm{FP}}$ and $\textrm{Recall} = \frac{\textrm{TP}}{\textrm{TP} + \textrm{FN}}$, where TP, FP and FN stand for true positives, false positives and false negatives. Due to the 10-fold stratified shuffle split used for model evaluation, the final result is an average F1-score across all 10 splits.


\section{Results}
\label{sec:evaluation}

In this section, we evaluate the performance of the proposed method against the state of the art methods from  \cite{bertalanivc2022resource}.  Table~\ref{tab:results-classification} compares our proposed method with three different deep learning models trained on time series transformed to images using recurrence plots (RP), Grammian Angular Summation Field (GASF) and Grammian Angular Differential Field (GADF). The first column displays the target classes, while the second column lists our proposed method using the three selected metrics. The remaining three columns show the results of the state of the art models. The best performing models for each class are bolded in the corresponding column.

As it can be seen in the first line of the binary classifier results in Table~\ref{tab:results-classification} the RP model achieves near perfect F1 score of 0.99 in detecting anomalous links. This is slightly better than the 0.98 F1 score of GASF model, while GADF model performs the worst out of the three with the F1 score of 0.91. Our proposed method has an F1 score lower by 0.05 compared to the RP model and 0.04 compared to GASF model. Looking at the results for GADF model, our method outperforms it with an F1 score higher by 0.03.

Looking at the second line of binary classifier results in Table~\ref{tab:results-classification} it can be seen that our proposed method outperforms all three state-of-the-art methods. Both the RP and GADF models are in terms of an F1 score outperformed by 0.01. Compared to GASF model our method achieves an F1 score higher by 0.04.

Although we compare our proposed method to the work from \cite{bertalanivc2022resource}, there is one main difference between the works. In \cite{bertalanivc2022resource} researchers were trying to detect whether the whole link is anomalous or not, while in our work we classify each measurement separately. The advantage of our approach is that not only can we detect whether an anomaly occurred in the link, we can also determine its location and duration within the link. Based on the location and duration we can then also determine what type of an anomaly occurred in the link. According to Table~\ref{tab:num-weights}, another benefit of our method is that our model has an $\approx$171 times less weights for slightly worse performance, compared to the state of the art imaging models.

\section{Conclusion}
\label{sec:conclusion}
In this paper, we performed a first time analysis of graph representation for wireless link layer anomalies detection on per TS point granularity. Additionally, we proposed a new graph neural network architecture that is able to detect and distinguish between the anomalous and non-anomalous points within the TS based on Graph Attention Networks. We compared our model to the existing state of the art. Our results show that our model in general performs similar to the state of the art deep learning model on image representation of time series, the main difference being that our method can localise which points within the time series are anomalous, while the state of can only detect whether an anomaly is present within the link, but can not provide an information on its location.

\section*{Acknowledgments}
This work was supported by the Slovenian Research Agency under grant P2-0016 and from the European Union’s Horizon Europe Framework Programme under grant agreement No 101096456.

\bibliographystyle{IEEEtran}
\bibliography{./references}

\begin{thebibliography}{10}
\providecommand{\url}[1]{#1}
\csname url@samestyle\endcsname
\providecommand{\newblock}{\relax}
\providecommand{\bibinfo}[2]{#2}
\providecommand{\BIBentrySTDinterwordspacing}{\spaceskip=0pt\relax}
\providecommand{\BIBentryALTinterwordstretchfactor}{4}
\providecommand{\BIBentryALTinterwordspacing}{\spaceskip=\fontdimen2\font plus
\BIBentryALTinterwordstretchfactor\fontdimen3\font minus
  \fontdimen4\font\relax}
\providecommand{\BIBforeignlanguage}[2]{{%
\expandafter\ifx\csname l@#1\endcsname\relax
\typeout{** WARNING: IEEEtran.bst: No hyphenation pattern has been}%
\typeout{** loaded for the language `#1'. Using the pattern for}%
\typeout{** the default language instead.}%
\else
\language=\csname l@#1\endcsname
\fi
#2}}
\providecommand{\BIBdecl}{\relax}
\BIBdecl

\bibitem{annaswamy2016emerging}
A.~M. Annaswamy, A.~R. Malekpour, and S.~Baros, ``Emerging research topics in
  control for smart infrastructures,'' \emph{Annual Reviews in Control},
  vol.~42, pp. 259--270, 2016.

\bibitem{cerar2020anomaly}
G.~Cerar, H.~Yetgin, B.~Bertalanic, and C.~Fortuna, ``Learning to detect
  anomalous wireless links in iot networks,'' \emph{IEEE Access}, vol.~8, pp.
  212\,130--212\,155, 2020.

\bibitem{silva2019m4dn}
J.~D.~C. Silva, J.~J.~P. Rodrigues, K.~Saleem, S.~A. Kozlov, and R.~A.
  Rab{\^e}lo, ``{M4DN. IoT-A Networks and Devices Management Platform for
  Internet of Things},'' \emph{IEEE Access}, vol.~7, pp. 53\,305--53\,313,
  April 2019.

\bibitem{sheth2006mojo}
A.~Sheth, C.~Doerr, D.~Grunwald, R.~Han, and D.~Sicker, ``Mojo: A distributed
  physical layer anomaly detection system for 802.11 wlans,'' in
  \emph{Proceedings of the 4th international conference on Mobile systems,
  applications and services}.\hskip 1em plus 0.5em minus 0.4em\relax ACM, 2006,
  pp. 191--204.

\bibitem{bertalanivc2022resource}
B.~Bertalani{\v{c}}, M.~Me{\v{z}}a, and C.~Fortuna, ``Resource-aware time
  series imaging classification for wireless link layer anomalies,'' \emph{IEEE
  Transactions on Neural Networks and Learning Systems}, 2022.

\bibitem{wu2020comprehensive}
Z.~Wu, S.~Pan, F.~Chen, G.~Long, C.~Zhang, and S.~Y. Philip, ``A comprehensive
  survey on graph neural networks,'' \emph{IEEE transactions on neural networks
  and learning systems}, vol.~32, no.~1, pp. 4--24, 2020.

\bibitem{zhou2020graph}
J.~Zhou, G.~Cui, S.~Hu, Z.~Zhang, C.~Yang, Z.~Liu, L.~Wang, C.~Li, and M.~Sun,
  ``Graph neural networks: A review of methods and applications,'' \emph{AI
  Open}, vol.~1, pp. 57--81, 2020.

\bibitem{ScheinertGNNanomalyCloud}
D.~Scheinert and A.~Acker, ``Telesto: A graph neural network model for anomaly
  classification in cloud services,'' in \emph{Service-Oriented Computing --
  ICSOC 2020 Workshops}, H.~Hacid, F.~Outay, H.-y. Paik, A.~Alloum,
  M.~Petrocchi, M.~R. Bouadjenek, A.~Beheshti, X.~Liu, and A.~Maaradji,
  Eds.\hskip 1em plus 0.5em minus 0.4em\relax Cham: Springer International
  Publishing, 2021, pp. 214--227.

\bibitem{deng2021graph}
A.~Deng and B.~Hooi, ``Graph neural network-based anomaly detection in
  multivariate time series,'' in \emph{Proceedings of the AAAI Conference on
  Artificial Intelligence}, vol.~35, no.~5, 2021, pp. 4027--4035.

\bibitem{blazWCNC}
B.~Bertalanic and C.~Fortuna, ``Graph isomorphism networks for wireless link
  layer anomaly classification,'' in \emph{2023 Wireless Communications and
  Networking Conference)}, 2023.

\bibitem{cao2020spectral}
D.~Cao, Y.~Wang, J.~Duan, C.~Zhang, X.~Zhu, C.~Huang, Y.~Tong, B.~Xu, J.~Bai,
  J.~Tong \emph{et~al.}, ``Spectral temporal graph neural network for
  multivariate time-series forecasting,'' \emph{Advances in neural information
  processing systems}, vol.~33, pp. 17\,766--17\,778, 2020.

\bibitem{wu2020connecting}
Z.~Wu, S.~Pan, G.~Long, J.~Jiang, X.~Chang, and C.~Zhang, ``Connecting the
  dots: Multivariate time series forecasting with graph neural networks,'' in
  \emph{Proceedings of the 26th ACM SIGKDD international conference on
  knowledge discovery \& data mining}, 2020, pp. 753--763.

\bibitem{Xuan_9695244}
Q.~Xuan, J.~Zhou, K.~Qiu, Z.~Chen, D.~Xu, S.~Zheng, and X.~Yang, ``Avgnet:
  Adaptive visibility graph neural network and its application in modulation
  classification,'' \emph{IEEE Transactions on Network Science and
  Engineering}, vol.~9, no.~3, pp. 1516--1526, 2022.

\bibitem{Xiu_9778887}
Y.~Xiu, X.~Ren, T.~Zhang, Y.~Chen, L.~Jiang, D.~Li, X.~Wang, L.~Zhao, and W.~K.
  Chan, ``Time labeled visibility graph for privacy-preserved physiological
  time series classification,'' in \emph{2022 7th International Conference on
  Cloud Computing and Big Data Analytics (ICCCBDA)}, 2022, pp. 280--284.

\bibitem{chandola2009anomaly}
V.~Chandola, A.~Banerjee, and V.~Kumar, ``Anomaly detection: A survey,''
  \emph{{ACM computing surveys (CSUR)}}, vol.~41, no.~3, July 2009.

\bibitem{salem2014anomaly}
O.~Salem, A.~Guerassimov, A.~Mehaoua, A.~Marcus, and B.~Furht, ``Anomaly
  detection in medical wireless sensor networks using svm and linear regression
  models,'' \emph{International Journal of E-Health and Medical Communications
  (IJEHMC)}, vol.~5, no.~1, pp. 20--45, 2014.

\bibitem{wazid2016efficient}
M.~Wazid and A.~K. Das, ``An efficient hybrid anomaly detection scheme using
  k-means clustering for wireless sensor networks,'' \emph{Wireless Personal
  Communications}, vol.~90, no.~4, pp. 1971--2000, 2016.

\bibitem{ahmad2019hybrid}
B.~Ahmad, W.~Jian, Z.~A. Ali, S.~Tanvir, and M.~S.~A. Khan, ``Hybrid anomaly
  detection by using clustering for wireless sensor network,'' \emph{Wireless
  Personal Communications}, vol. 106, no.~4, pp. 1841--1853, 2019.

\bibitem{bertalanic_9606264}
B.~Bertalanic, H.~Yetgin, G.~Cerar, and C.~Fortuna, ``A deep learning model for
  anomalous wireless link detection,'' in \emph{2021 17th International
  Conference on Wireless and Mobile Computing, Networking and Communications
  (WiMob)}, 2021, pp. 265--270.

\bibitem{thing2017ieee}
V.~L. Thing, ``{IEEE 802.11} network anomaly detection and attack
  classification: {A} deep learning approach,'' in \emph{IEEE Wireless
  Communications and Networking Conference {(WCNC)}}, San Francisco, CA, USA,
  March 2017.

\bibitem{ran2019semi}
J.~Ran, Y.~Ji, and B.~Tang, ``A semi-supervised learning approach to ieee
  802.11 network anomaly detection,'' in \emph{2019 IEEE 89th Vehicular
  Technology Conference (VTC2019-Spring)}.\hskip 1em plus 0.5em minus
  0.4em\relax IEEE, 2019, pp. 1--5.

\bibitem{wang2015encoding}
Z.~Wang and T.~Oates, ``Encoding time series as images for visual inspection
  and classification using tiled convolutional neural networks,'' in
  \emph{Workshops at the twenty-ninth AAAI conference on artificial
  intelligence}, 2015.

\bibitem{velickovic2018graph}
\BIBentryALTinterwordspacing
P.~Veli{\v{c}}kovi{\'{c}}, G.~Cucurull, A.~Casanova, A.~Romero, P.~Li{\`{o}},
  and Y.~Bengio, ``{Graph Attention Networks},'' \emph{International Conference
  on Learning Representations}, 2018. [Online]. Available:
  \url{https://openreview.net/forum?id=rJXMpikCZ}
\BIBentrySTDinterwordspacing

\bibitem{skip_connection}
S.~A. Taghanaki, A.~Bentaieb, A.~Sharma, S.~K. Zhou, Y.~Zheng, B.~Georgescu,
  P.~Sharma, Z.~Xu, D.~Comaniciu, and G.~Hamarneh, ``Select, attend, and
  transfer: Light, learnable skip connections,'' in \emph{Machine Learning in
  Medical Imaging}, H.-I. Suk, M.~Liu, P.~Yan, and C.~Lian, Eds.\hskip 1em plus
  0.5em minus 0.4em\relax Cham: Springer International Publishing, 2019, pp.
  417--425.

\bibitem{rutgers-noise-20070420}
S.~K. Kaul, I.~Seskar, and M.~Gruteser, ``{CRAWDAD} dataset rutgers/noise (v.
  2007-04-20),'' Downloaded from
  \url{https://crawdad.org/rutgers/noise/20070420/RSSI}, Apr. 2007, traceset:
  RSSI.

\end{thebibliography}

\end{document}